\documentclass[10pt]{article}
\usepackage{geometry}
\usepackage{xcolor}
\definecolor{graycolor}{gray}{0.9}
\usepackage{microtype}
\usepackage{setspace}
\usepackage{soul}
\usepackage{cmap}
\usepackage[utf8]{inputenc}
\usepackage[T1]{fontenc}
\usepackage[english]{babel}
\usepackage{times}
\usepackage[numbers,sort&compress]{natbib}

\usepackage{titlesec}
\titleformat {\section} [block] {\raggedright \fontsize{10}{10}\selectfont\bfseries} {\thesection. \space} {0pt} {}
\titlespacing {\section} {0pt} {12pt} {6pt}
\titleformat {\subsection} [block] {\raggedright \fontsize{10}{10}\selectfont\itshape} {\thesubsection .\space} {0pt} {}
\titlespacing {\subsection} {0pt} {12pt} {6pt}
\titleformat {\subsubsection} [block] {\raggedright \fontsize{10}{10}\selectfont} {\thesubsubsection .\space} {0pt} {}
\titlespacing {\subsubsection} {0pt} {12pt} {6pt}
\titleformat {\paragraph} [block] {\raggedright \fontsize{10}{10}\selectfont} {} {0pt} {}
\titlespacing {\paragraph} {0pt} {12pt} {6pt}

\usepackage{array}
\newcommand{\PreserveBackslash}[1]{\let\temp=\\#1\let\\=\temp}
\newcolumntype{C}[1]{>{\PreserveBackslash\centering}m{#1}}
\newcolumntype{R}[1]{>{\PreserveBackslash\raggedleft}m{#1}}
\newcolumntype{L}[1]{>{\PreserveBackslash\raggedright}m{#1}}
\usepackage{lineno}
\usepackage{tabularx}
\usepackage{colortbl}
\usepackage{graphicx}
\usepackage{float}
\usepackage[export]{adjustbox}

\usepackage{caption}
\usepackage{fancyhdr}
\usepackage{amsmath}
\usepackage{lastpage}
\usepackage{layout}
\usepackage{hyperref}
\hypersetup{
	colorlinks=true,
	linkcolor=blue,
	filecolor=blue,
	urlcolor=black,
	citecolor=cyan,
}

\setcitestyle{open={[},close={]},citesep={,\!},numbers}

\fancypagestyle{firstpage}{
    \setlength{\headsep}{2.2cm}
    
    \setlength{\footskip}{1.5cm}
    \fancyhf{}% clear default for head and foot
    \lhead{\begin{table}[H]
        \centering
        \begin{tabular}{L{2.5cm}C{10cm}C{3.1cm}R{2cm}}
            \includegraphics[scale=0.035]{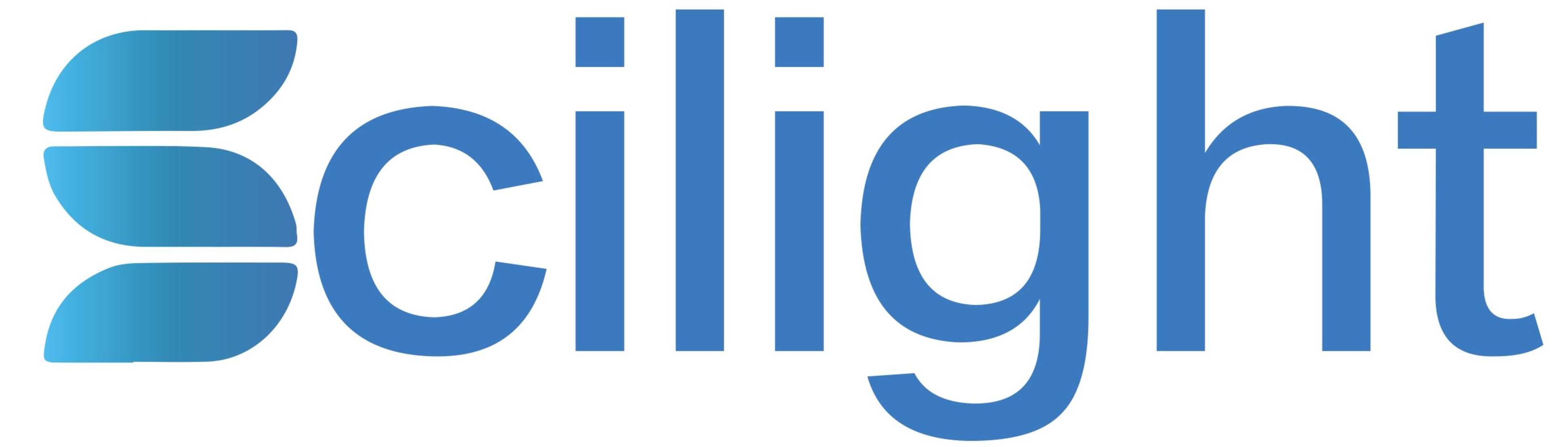}
            \vspace{-6pt}
            & \cellcolor{graycolor}
            \begin{tabular}[c]{@{}c@{}}
            \textit{International Journal of Gravitation and Theoretical Physics}\\ \href{https://www.sciltp.com/journals/ijgtp}{https://www.sciltp.com/journals/ijgtp}
            \end{tabular}
            &
            \includegraphics[scale=0.047]{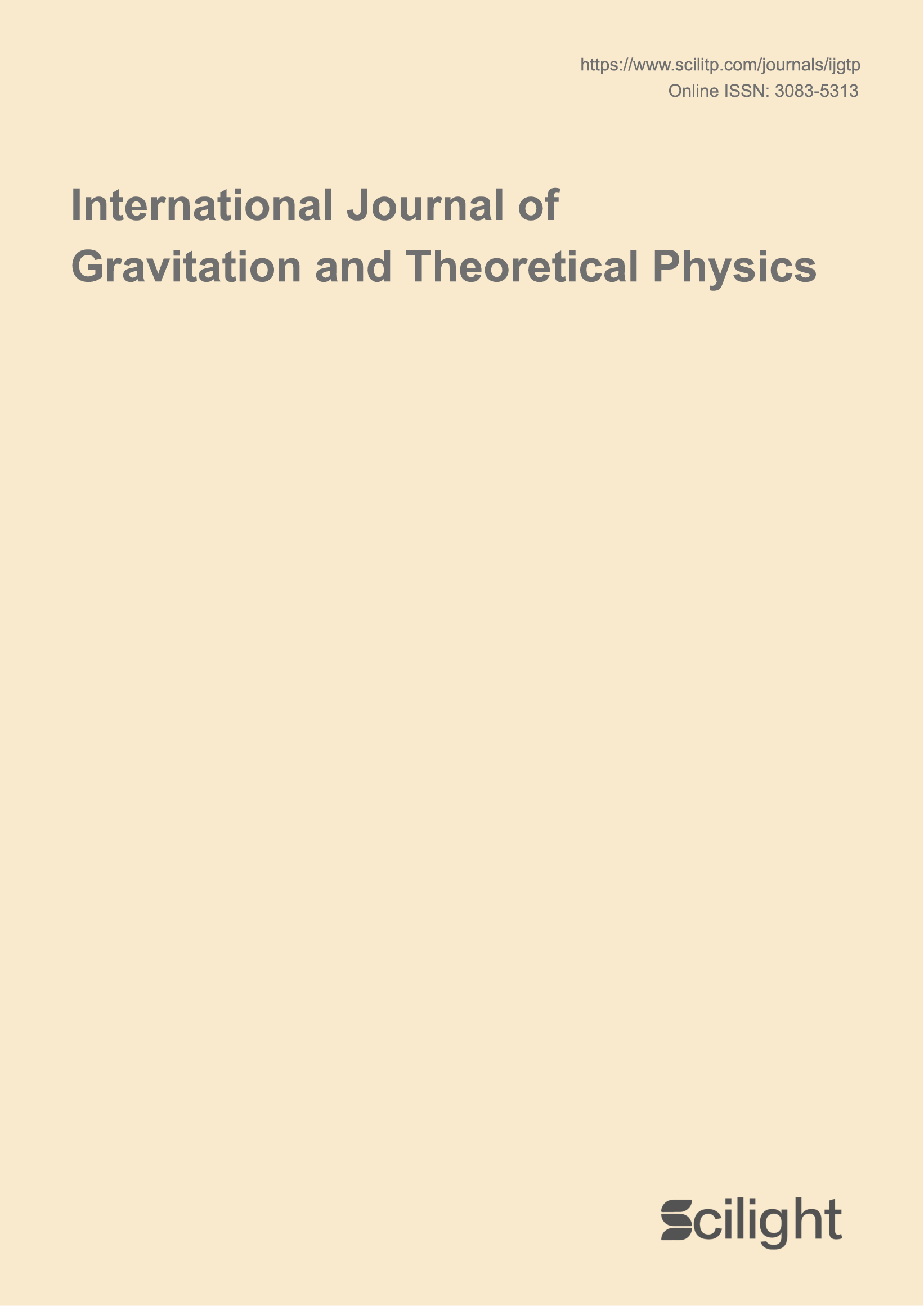}
            \vspace{-3pt}\\
        \end{tabular}
        \vspace{-22pt}
    \end{table}}
   
    \fancyfoot[C]{
        \vspace{-1.55cm}
        \begin{table}[H]
            \begin{minipage}[c]{0.15\columnwidth}
                \includegraphics[scale=0.67]{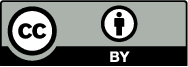}
                \vspace{1.1pt}
            \end{minipage}
            \hfill
            \begin{minipage}[c]{0.85\columnwidth}
                \scriptsize \textbf{Copyright:} © 2026 by the authors. This is an open access article under the terms and conditions of the Creative Commons Attribution (\mbox{CC BY}) license (\href{https://creativecommons.org/licenses/by/4.0/}{https://creativecommons.org/licenses/by/4.0/}). \\ \textbf{Publisher’s Note:} Scilight stays neutral with regard to jurisdictional claims in published maps and institutional affiliations.
            \end{minipage}
    \end{table}}
    \vspace{-0.55cm}
}

\def\imo{i}
\def\Order#1{{\cal O}\left(#1\right)}
\def\re#1{Re(#1)}
\def\im#1{Im(#1)}

\begin{document}
\newgeometry{left=2.5cm, right=2.5cm, top=1.8cm, bottom=4cm}
	\thispagestyle{firstpage}
	\nolinenumbers
	{\noindent \textit{Article}}
	\vspace{4pt} \\
	{\fontsize{18pt}{10pt}\textbf{An efficient higher-order WKB code \\
    for quasinormal modes and greybody factors}  }
	\vspace{16pt} \\
	{\large Roman. A. Konoplya \textsuperscript{1,*}, Jerzy Matyjasek\textsuperscript{2}, and Alexander Zhidenko \textsuperscript{1,3} }
	\vspace{6pt}
	\begin{spacing}{0.9}
		{\noindent \small
			\textsuperscript{1}	Research Centre for Theoretical Physics and Astrophysics, Institute of Physics, Silesian University in Opava, \\ \hspace*{0.35em} Bezručovo náměstí 13, CZ-74601 Opava, Czech Republic \\
            \textsuperscript{2} Institute of Physics, Maria Curie-Skłodowska University, pl. Marii Curie-Skłodowskiej 1, 20-031 Lublin, Poland, \\
			\textsuperscript{3}	Centro de Matemática, Computação e Cognição (CMCC), Universidade Federal do ABC (UFABC), \\
            \hspace*{0.4em} Rua Abolição, Santo André 09210-180, SP, Brazil \\
		    {*}  \parbox[t]{0.98\linewidth}{Correspondence: roman.konoplya@gmail.com}
            \vspace{6pt}\\
		    \footnotesize \textbf{How To Cite}: Konoplya, R.A.; Matyjasek, J; Zhidenko, A.; An efficient higher-order WKB code for quasinormal modes and greybody factors. \emph{International Journal of Gravitation and Theoretical Physics} \textbf{2026}, \emph{2}(1), 5. \href{https://doi.org/10.53941/ijgtp.2026.100005}{https://doi.org/10.53941/ijgtp.2026.100005} \href{https://arxiv.org/abs/2603.12466}{[arXiv:2603.12466]}
        }\\
	\end{spacing}

\begin{table}[H]
\noindent\rule[0.15\baselineskip]{\textwidth}{0.5pt}
\begin{tabular}{lp{12.4cm}}
 \small
  \begin{tabular}[t]{@{}l@{}}
  \footnotesize  Received: 8 March 2026 \\
  \footnotesize  Revised: 17 March 2026 \\
  \footnotesize  Accepted: 19 March 2026 \\
  \footnotesize  Published: 23 March 2026
  \end{tabular} &
  \textbf{Abstract:} The higher-order WKB Mathematica® code for computing quasinormal modes, whose accuracy was significantly enhanced through extensions to higher orders and, in particular, through the use of Padé resummation, has been widely employed in numerous studies over the past several years. In this work, we present an updated and optimized version of the code. The main improvement consists in expanding the effective potential in a Taylor series around its maximum, rather than evaluating the full analytic expression of the WKB formula for each specific potential. This modification leads to a substantial reduction in computation time. In cases where the effective potential is complicated and involves non-rational functions, the speed gain can reach several orders of magnitude, while preserving the accuracy of the method.
  \\
  \\
  &
  &

\vspace{4pt}


\end{tabular}
\noindent\rule[0.15\baselineskip]{\textwidth}{0.5pt}
\end{table}

\section{Introduction}\label{sec:introduction}

Quasinormal modes and grey-body factors are two fundamental characteristics describing wave dynamics in black-hole spacetimes \cite{Kokkotas:1999bd,Konoplya:2011qq,Berti:2009kk,Bolokhov:2025uxz,Page:1976df,Futterman:1988ni}. The quasinormal spectrum governs the ringdown phase of perturbations and encodes information about the geometry and stability of the background, playing a central role in gravitational-wave phenomenology  \cite{LIGOScientific:2016aoc,LIGOScientific:2017vwq,LIGOScientific:2020zkf,Babak:2017tow}. Grey-body factors, on the other hand, determine the transmission probability of waves through the effective potential barrier and thus control the spectrum of Hawking radiation and scattering properties of the black hole. Together, quasinormal modes and grey-body factors provide complementary insights into the dynamical and radiative aspects of black-hole physics, linking classical perturbation theory with quantum emission processes.

An efficient and economical approach for evaluating both quasinormal modes and grey-body factors is provided by the Wentzel--Kramers--Brillouin (WKB) approximation. Originally introduced into black-hole perturbation theory by Schutz, Iyer and Will \cite{Schutz:1985km,Iyer:1986np}, the method exploits the barrier-like structure of the effective potential and allows one to extract characteristic frequencies and transmission coefficients from local information near its maximum. Owing to its simplicity and relatively low computational cost, the WKB approach has become a standard and widely used tool in studies of black-hole spectroscopy and wave scattering \cite{Sekhmani:2025jbl,Ou:2025bbv,Lutfuoglu:2025qkt,Bolokhov:2025egl,Deng:2025hfn,Bambagiotti:2025qxj,Feng:2025iao,AraujoFilho:2025vgb,Liu:2025iby,Aoki:2025ihc,Bohra:2025oro,Arbelaez:2025gwj,Lutfuoglu:2025pzi,Hamil:2025fbn,Pathrikar:2025gzu,Sajjad:2025xdo,Momennia:2025vkv,Spina:2025wxb,Liang:2025rbe,Pathrikar:2025sin,Malik:2025czt,AraujoFilho:2025zaj,Bolokhov:2025fto,Sekhmani:2025yjf,Bolokhov:2025zva,dePaiva:2025eux,Tang:2025qaq,Liang:2026eic,Pereira:2026ffn,Qi:2026zrr,Lutfuoglu:2026xlo}.

While the previously released public Mathematica® implementation of the WKB method \cite{Konoplya:2019hlu} incorporated corrections up to the 13th order together with Padé resummation \cite{Matyjasek:2017psv}, the present version extends the expansion up to the 16th order \cite{Matyjasek:2019eeu}. More importantly, the new code evaluates the derivatives of the effective potential numerically at its maximum, rather than constructing full analytic expressions for the higher-order derivatives. In cases where the potential is algebraically complicated or contains non-rational functions, the symbolic differentiation required by the earlier implementation becomes extremely time-consuming and may even render the computation impractical. By relying on the values of derivatives at the peak, the updated code achieves a dramatic improvement in performance: calculations that previously required hours, or were effectively unfeasible, are now completed within a fraction of a second.

The manuscript is organized as follows. In Sec.~\ref{sec:WKB}, we briefly outline the WKB approach for the calculation of quasinormal modes and grey-body factors. Section~\ref{sec:improvements} summarizes the improvements implemented in the Mathematica® package. In Sec.~\ref{sec:expansion}, we present analytic expansions beyond the eikonal approximation for quasinormal modes and grey-body factors derived from the WKB formula, while Sec.~\ref{sec:correspondence} is devoted to the correspondence between quasinormal frequencies and grey-body factors, which is also incorporated into the package. Finally, in Sec.~\ref{sec:conclusions}, we summarize the main features of the package and provide an outlook on possible extensions and future applications of the presented methods. In addition, an ancillary Mathematica® notebook containing illustrative examples of the discussed WKB applications is made available at \url{https://arxiv.org/src/2603.12466v1/anc}.

\section{Boundary conditions and essentials of the WKB approach}\label{sec:WKB}

We consider the general scattering problem for a wavelike equation
\begin{equation}\label{wavelike}
\frac{d^2\Psi}{dx^2} = U(x,\omega)\Psi ,
\end{equation}
where the effective potential $U(x,\omega)$ depends on a nonvanishing frequency $\omega\neq0$\footnote{Static solutions are not considered.} and has the form of a single-barrier potential, approaching negative constants as $x\to\pm\infty$. When the effective potential is asymptotically constant, solutions of Eq.~\eqref{wavelike} in the asymptotic regions are superpositions of ingoing and outgoing waves.

We assume the time dependence of the perturbation function to be proportional to $\propto e^{-\imo\omega t}$.
The scattering problem is defined by the boundary conditions
\begin{equation}\label{BC}
\begin{array}{rcll}
\Psi &=& \Psi^+_{\text{in}}(x) + R\,\Psi^+_{\text{out}}(x), & x \to +\infty, \\
\Psi &=& T\,\Psi^-(x), & x \to -\infty,
\end{array}
\end{equation}
where $R$ and $T$ are the reflection and transmission coefficients, respectively.
Here $\Psi^-$ is the wave going to the left asymptotic, satisfying,
\begin{equation}\label{ingoinghorizon}
\Psi^-(x \to -\infty) \propto
\left\{
\begin{array}{ll}
e^{- \imo k_- x}, & \omega > 0, \\
e^{ \imo k_- x}, & \omega < 0,
\end{array}
\right.
\end{equation}
while $\Psi^+_{\text{in}}(x)$ and $\Psi^+_{\text{out}}(x)$ are, respectively, the ingoing and outgoing wave at the right asymptotic, such that
\begin{equation}\label{ingoingwave}
\Psi^+_{\text{in}}(x \to \infty) \propto
\left\{
\begin{array}{ll}
e^{- \imo k_+ x}, & \omega > 0, \\
e^{ \imo k_+ x}, & \omega < 0,
\end{array}
\right.
\end{equation}
and
\begin{equation}\label{outgoingwave}
\Psi^+_{\text{out}}(x \to \infty) \propto
\left\{
\begin{array}{ll}
e^{ \imo k_+ x}, & \omega > 0, \\
e^{- \imo k_+ x}, & \omega < 0 .
\end{array}
\right.
\end{equation}

The asymptotic wave numbers satisfy the dispersion relations
\begin{equation}
k_\pm^2(\omega) = - \lim_{x \to \pm\infty} U(x,\omega),
\end{equation}
and we choose $\re{k_\pm}>0$.

In black-hole perturbation theory, Eq.~\eqref{wavelike} is typically written in terms of the tortoise coordinate,
\begin{equation}\label{tortoise}
  dx\equiv\frac{dr}{f(r)},
\end{equation}
where $r$ is the radial coordinate and $f(r)$ is the blackening factor, so that $x\to+\infty$ corresponds to spatial infinity and $x\to-\infty$ to the event horizon.

In order to obtain the reflection and transmission coefficients, we employ the WKB approximation. The WKB method is based on matching the asymptotic solutions with the Taylor expansion of the solution near the maximum of the potential barrier at $x=x_m$. The matching is performed through the two turning points, defined as the zeros of $U(x,\omega)$. As a result, one obtains the S-matrix relating the asymptotic amplitudes, expressed in terms of the value of the potential at its maximum,
\begin{equation}
U_0(\omega) \equiv U(x_m,\omega),
\end{equation}
and its higher derivatives,
\begin{equation}
U_1(\omega) = \left. \frac{d U}{dx} \right|_{x=x_m}=0, \quad U_2(\omega) = \left. \frac{d^2 U}{dx^2} \right|_{x=x_m}, \quad
U_3(\omega) = \left. \frac{d^3 U}{dx^3} \right|_{x=x_m}, \quad \ldots.
\end{equation}

The resulting WKB condition takes the form
\begin{equation}\label{WKBformula}
0 = U_0(\omega) + A_2(\mathcal{K}^2) + A_4(\mathcal{K}^2) + A_6(\mathcal{K}^2) + \ldots  - \imo \mathcal{K} \sqrt{-2 U_2(\omega)}
\left( 1 + A_3(\mathcal{K}^2) + A_5(\mathcal{K}^2) + A_7(\mathcal{K}^2) + \ldots \right),
\end{equation}
where $A_k(\mathcal{K}^2)$ are the WKB corrections of order $k$. These corrections are polynomials in $\mathcal{K}^2$ and the derivatives
$U_2(\omega), U_3(\omega),\ldots, U_{2k}(\omega)$ with rational coefficients, divided by appropriate powers of $U_2(\omega)$. The explicit forms of $A_2(\mathcal{K}^2)$ and $A_3(\mathcal{K}^2)$ were obtained in \cite{Iyer:1986np}, the corrections $A_4(\mathcal{K}^2)$, $A_5(\mathcal{K}^2)$, and $A_6(\mathcal{K}^2)$ were derived in \cite{Konoplya:2003ii}, and the higher-order terms were computed in \cite{Matyjasek:2017psv,Matyjasek:2019eeu}.

For a real effective potential, the quantity $\mathcal{K}$ is purely imaginary and is related to the reflection and transmission coefficients by \cite{Iyer:1986np}
\begin{eqnarray}
|R|^2 &=& \frac{1}{1 + e^{-2\pi \imo \mathcal{K}}}, \qquad 0 < |R|^2 < 1, \label{reflection} \\
|T|^2 &=& \frac{1}{1 + e^{2\pi \imo \mathcal{K}}} = 1 - |R|^2. \label{transmission}
\end{eqnarray}
The transmission probability defines the grey-body factor,
\begin{equation}\label{greybodyfactor}
\Gamma(\omega)\equiv |T|^2 = \frac{1}{1 + e^{2\pi \imo \mathcal{K}}}.
\end{equation}

At the eikonal level, Eq.~\eqref{WKBformula} yields the unique solution
\begin{equation}\label{eikonalK}
\mathcal{K} = - \imo\frac{U_0(\omega)}{\sqrt{-2 U_2(\omega)}} ,
\end{equation}
while higher-order WKB equations possess multiple roots for a given $\omega$. As long as we consider the terms $A_2(\mathcal{K}^2),A_3(\mathcal{K}^2),\ldots$ as small corrections to the eikonal approximation, the physically relevant root to Eq.~\eqref{WKBformula} can be chosen as the one closest to the eikonal value.

Quasinormal modes can be formulated as a special case of a scattering problem, in which the corresponding solutions are purely outgoing at spatial infinity and purely ingoing at the event horizon. They are characterized by complex eigenfrequencies $\omega$, whose real parts represent the oscillation frequencies, while the negative imaginary parts determine the decay rates. Within the WKB approach, quasinormal frequencies are obtained via analytic continuation of the S-matrix into the complex frequency plane and are associated with the poles of the Gamma functions \cite{Iyer:1986np}
$$\Gamma\left(-\mathcal{K}+\frac{1}{2}\right) \quad\mbox{for}\quad \re{\omega}>0 \qquad\mbox{and}\qquad \Gamma\left(\mathcal{K}+\frac{1}{2}\right) \quad\mbox{for}\quad \re{\omega}<0.$$
This leads to the quantization condition
\begin{align}\label{QNMsK}
\mathcal{K} =
\left\{
\begin{array}{cc}
+n + \tfrac{1}{2}, & \re{\omega} > 0, \\
-n - \tfrac{1}{2}, & \re{\omega} < 0,
\end{array}
\right.
\qquad
n = 0,1,2,\ldots.
\end{align}
For these values of $\mathcal{K}$, the denominators in Eqs.~\eqref{reflection} and \eqref{transmission} vanish, reflecting the fact that quasinormal modes correspond to poles of the reflection and transmission coefficients.

In practice, quasinormal frequencies are computed by fixing the parameters of the effective potential, determining numerically the position $x_m(\omega)$ of its maximum, and solving Eq.~\eqref{WKBformula} for $\mathcal{K}$ given by Eq.~\eqref{QNMsK}. In many spherically symmetric black-hole spacetimes, the effective potential depends on the frequency in the simple form
\begin{equation}
U(x,\omega) = V(x) - \omega^2.
\end{equation}

In this case, Eq.~\eqref{WKBformula} gives a closed-form expression for the quasinormal-mode frequencies $\omega_n$ as functions of the overtone number $n$ and the derivatives of the effective potential $V(x)$ evaluated at its maximum. As proposed in \cite{Matyjasek:2017psv}, Padé rational approximants can be applied to the WKB formula, interpreted as a polynomial expansion in higher-order correction terms. Balanced Padé approximants are found to provide the most stable and reliable results, yielding accurate approximations for the dominant quasinormal frequencies in the regime $|\re{\omega_n}|\geq|\im{\omega_n}|$.

It was argued in \cite{Hatsuda:2019eoj} that the WKB expansion for quasinormal frequencies exhibits properties consistent with Borel summability. In this approach, applying Borel summation to the formal WKB series allows one to reconstruct highly accurate values of the quasinormal frequencies, indicating that the asymptotic WKB expansion captures the relevant physical information about the spectrum despite its divergent nature. Higher-order WKB corrections can be obtained numerically by mapping the quasinormal-mode problem to the bound-state spectrum of an anharmonic oscillator and applying Rayleigh–Schrödinger perturbation theory around the maximum of the potential \cite{Hatsuda:2019eoj,Hatsuda:2021gtn}. This formulation enables the implementation of the WKB method to very high orders and can yield highly accurate results when combined with Borel summation. However, in practice the numerical implementation of the Borel summation may be demanding. Therefore, in the present review we focus primarily on the WKB approach with Padé resummation, which has been shown to be numerically stable and produce results of comparable accuracy \cite{Matyjasek:2019eeu}.

\section{Improvements in the package}\label{sec:improvements}

Here we summarize the main improvements of the package introduced in Ref.~\cite{Konoplya:2019hlu}. In addition to the explicit implementation of the sixteenth-order WKB formula derived in Ref.~\cite{Matyjasek:2019eeu}, the package employs numerical series expansions of the effective potential $V(r)$ and the blackening factor $f(r)$ around the (numerically determined) maximum of the effective potential, rather than relying on analytic computations of higher-order derivatives with respect to the tortoise coordinate followed by substitution of the potential maximum. As a result, the computation of quasinormal frequencies and reflection/transmission coefficients is significantly accelerated, while numerical precision loss associated with floating-point operations is substantially reduced.

\begin{table}
\begin{tabular}{|r||rr|rr|rr||rr|rr|rr||}
\multicolumn{7}{c}{Schwarzschild-de Sitter (Dirac field)} & \multicolumn{6}{c}{Proper time flow regular black hole (scalar)}\\
\hline
&\multicolumn{4}{c|}{quasinormal mode} &\multicolumn{2}{c||}{$\mathcal{K}$}& \multicolumn{4}{c|}{quasinormal mode} &\multicolumn{2}{c||}{$\mathcal{K}$}\\
\!WKB\! & \multicolumn{2}{c|}{~time (seconds)~} & \multicolumn{2}{c|}{loss (places)} & \multicolumn{2}{c||}{~time (seconds)~} & \multicolumn{2}{c|}{~time (seconds)~} & \multicolumn{2}{c|}{loss (places)} & \multicolumn{2}{c||}{~time (seconds)~} \\
\!order\! & old & new & old & new & old & new & old & new & old & new & old & new \\
\hline
  1 & 0.059 & 0.0028 &   7 &  9 &  0.109 & 0.032 &     1 &    0.0284 &   8 & 12 &     1 & 0.047 \\
  2 & 0.147 & 0.0028 &  14 & 14 &  0.281 & 0.093 &    21 &    0.0325 &  64 & 17 &    25 & 0.109 \\
  3 & 0.303 & 0.0032 &  21 & 18 &  0.563 & 0.188 &    67 &    0.0344 & 125 & 22 &    67 & 0.172 \\
  4 & 0.491 & 0.0034 &  29 & 23 &  0.953 & 0.312 &   178 &    0.0350 & 185 & 27 &   220 & 0.453 \\
  5 & 0.744 & 0.0039 &  37 & 28 &  1.375 & 0.391 &   418 &    0.0353 & 245 & 33 &   426 & 0.563 \\
  6 & 1.047 & 0.0048 &  45 & 34 &  2.016 & 0.641 &  1069 &    0.0372 & 306 & 39 &  1742 & 1.281 \\
  7 & 1.509 & 0.0066 &  53 & 40 &  2.875 & 0.796 &  8129 &    0.0388 & 350 & 47 &  8205 & 1.422 \\
  8 & 1.912 & 0.0114 &  61 & 46 &  4.547 & 0.922 &       &    0.0450 &     & 53 &       & 2.422 \\
  9 & 2.531 & 0.0177 &  69 & 52 &  8.718 & 1.563 &       &    0.0562 &     & 59 &       & 2.625 \\
 10 & 3.094 & 0.0306 &  78 & 58 & 11.125 & 1.750 &       &    0.0766 &     & 67 &       & 3.937 \\
 11 & 4.200 & 0.0510 &  87 & 64 & 16.797 & 2.750 &       &    0.1106 &     & 72 &       & 5.469 \\
 12 & 5.491 & 0.0966 &  95 & 71 & 16.438 & 3.062 &       &    0.1738 &     & 78 &       & 5.797 \\
 13 & 5.756 & 0.2412 & 104 & 77 & 32.281 & 3.672 &       &    0.2825 &     & 85 &       & 6.281 \\
\hline
\end{tabular}
\caption{Comparison of computation time (in seconds) and precision loss (in binary places) for the dominant mode as well as computation time of the S-matrix coefficient $\mathcal{K}$ for 100 values of $\omega$ of a test Dirac field in the Schwarzschild–de Sitter background ($M=1$, $\Lambda=0.01$) and a test scalar field in the background of a regular black hole from proper time flow in quantum gravity ($M=1$, $q=2$). The new package (right columns) is significantly faster and exhibits reduced precision loss from floating-point operations (except for the eikonal approximation). For more cumbersome effective potentials, the performance advantage becomes even more pronounced. All calculations were performed on an Intel® Core™ i5-1334U using Wolfram Mathematica® 7.0.0 (precision=200, \$MaxExtraPrecision=1000).}\label{tabl:efficiency}
\end{table}

A quantitative comparison of computation time and precision loss is presented in Table~\ref{tabl:efficiency} for quasinormal modes and grey-body factors of the massless Dirac field in the Schwarzschild–de Sitter background~\cite{Zhidenko:2003wq}, as well as for a test scalar field in the background of a regular black hole arising from proper-time flow in quantum gravity~\cite{Bonanno:2025dry}. In the latter case, the original package becomes impractical for computing higher-order WKB corrections even for scalar perturbations, owing to the cumbersome analytic structure of the metric function and the resulting complexity of the effective potential. By contrast, in the new package the computation time at higher WKB orders is nearly independent of the complexity of the effective potential. This improvement makes the method particularly suitable for complicated gravitational backgrounds, and the package has already been successfully applied to the calculation of quasinormal modes of black holes in quasi-topological gravity \cite{Konoplya:2026gim}. When computing S-matrix coefficients using the new package, the numerical solution of the polynomial equation with respect to $\mathcal{K}$ is, in fact, significantly more time-consuming than the derivation of the corresponding higher-order WKB formulas.

The location of the potential maximum may be determined numerically, specified analytically, or expressed through a series expansion in terms of arbitrary parameters. In particular, the expansion in inverse powers of the multipole number provides a systematic framework for deriving analytic expressions for quasinormal modes and grey-body factors beyond the standard eikonal approximation. This approach allows one to compute higher-order corrections in a controlled manner and to assess the accuracy and convergence of the WKB expansion in regimes where the leading-order eikonal formulas are insufficient. A detailed review of this method and its implementation is presented in the next section.

\section{Analytic expansion beyond eikonal approximation}\label{sec:expansion}

The eikonal approximation for quasinormal mode frequencies follows from the first-order WKB formula,
\begin{equation}
\omega=\sqrt{V_0-\imo \mathcal{K}\sqrt{-2V_2}},
\end{equation}
by expanding the result in inverse powers of the large parameter $\kappa^{-1}$,
\begin{equation}\label{eikonal-formula}
\omega=\Omega\kappa-\imo\lambda\mathcal{K}+\Order{\kappa^{-1}},
\end{equation}
where $\kappa\equiv\ell+\tfrac{1}{2}$ and $\ell=s,s+1,s+2,\ldots$ is the (half-)integer multipole number, whose minimal value is determined by the spin of the perturbation $s$. The expansion~\eqref{eikonal-formula} is valid for large $\ell$ and becomes exact in the asymptotic eikonal limit $\ell\to\infty$.

A systematic extension of this approximation beyond leading eikonal order was proposed in Ref.~\cite{Konoplya:2023moy}, based on higher-order WKB formulas. In this approach, the location of the maximum of the effective potential is expanded as a series in inverse powers of $\kappa$,
\begin{equation}\label{rmax}
x_m=x_0+x_1\kappa^{-1}+x_2\kappa^{-2}+\cdots+x_k\kappa^{-k}+\Order{\kappa}^{-k-1},
\end{equation}
and the resulting expression is substituted into the WKB formula \eqref{WKBformula} of the same order $k$. This procedure yields an analytic expansion of the quasinormal frequencies that consistently incorporates higher-order corrections beyond the eikonal limit.

It was demonstrated in Ref.~\cite{Konoplya:2023moy} that this method provides high accuracy even for the lowest allowed multipole numbers, including $\ell=s$, in the case of the Schwarzschild–de Sitter black hole. Owing to its robustness and efficiency, this approach has since been applied in a number of subsequent studies \cite{Bolokhov:2023bwm,Konoplya:2025mvj,Malik:2024qsz,Malik:2024itg,Malik:2024bmp,Malik:2024tuf,Malik:2024wvs}, confirming its usefulness well beyond the regime of large $\ell$.

For practical implementation, the appropriate routines for constructing the series expansion of the potential peak location~\eqref{rmax}, as well as for its efficient substitution into the higher-order WKB formula, are provided in the package. This allows the expansion to be generated to higher order in $\kappa^{-1}$ with minimal computational overhead. In addition, the method admits a straightforward generalization to simultaneous expansions in several small parameters, in addition to $\kappa^{-1}$. This makes it possible to derive analytic approximations for the quasinormal spectrum of parametrized families of black holes and to perform a qualitative analysis of the dependence of quasinormal modes on the underlying model parameters.

The resulting expansion can be inverted to obtain $\mathcal{K}$ as a series in inverse powers of $\kappa^{-1}$. Following Ref.~\cite{Konoplya:2023moy}, one assumes that
\begin{equation}
\omega^2-\Omega^2\kappa^2=\Order{\kappa},
\end{equation}
which is consistent with the eikonal behavior of the quasinormal spectrum and ensures the self-consistency of the expansion. Under this assumption, Eq.~\eqref{WKBformula} is solved perturbatively with respect to $\mathcal{K}$, yielding an explicit analytic approximation for the corresponding S-matrix coefficient. As a result, this approach allows for an analytic evaluation of the grey-body factors, providing insight into their dependence on the multipole number and on the parameters characterizing the black hole geometry.

\section{Correspondence between grey-body factors and quasinormal modes}\label{sec:correspondence}

Since the WKB formula~\eqref{sec:WKB} relates the frequency $\omega$ and the parameter $\mathcal{K}$ through the value of the effective potential and its derivatives at the potential peak, one can express analytic approximations for the S-matrix coefficients directly in terms of quasinormal-mode frequencies \cite{Konoplya:2024lir}.

Within the eikonal approximation, the S-matrix coefficient admits a particularly simple representation in terms of the dominant quasinormal frequency $\omega_0$,
\begin{equation}\label{eq:gbeikonal}
\imo\mathcal{K}=\frac{\omega^2-\re{\omega_0}^2}{4\re{\omega_0}\im{\omega_0}}+\Order{\frac{1}{\kappa}}.
\end{equation}

The approximation can be systematically improved by including higher-order WKB corrections. Retaining terms up to third order yields \cite{Konoplya:2024lir}
\begin{eqnarray}\label{eq:gbsecondorder}
\imo\mathcal{K}&=&\frac{\omega^2-\re{\omega_0}^2}{4\re{\omega_0}\im{\omega_0}}\left(1+\frac{(\re{\omega_0}-\re{\omega_1})^2}{32\im{\omega_0}^2}-\frac{3\im{\omega_0}-\im{\omega_1}}{24\im{\omega_0}}\right)
\\\nonumber&& -\frac{\re{\omega_0}-\re{\omega_1}}{16\im{\omega_0}}-\frac{(\omega^2-\re{\omega_0}^2)^2}{16\re{\omega_0}^3\im{\omega_0}}\left(1+\frac{\re{\omega_0}(\re{\omega_0}-\re{\omega_1})}{4\im{\omega_0}^2}\right)
\\\nonumber&& +\frac{(\omega^2-\re{\omega_0}^2)^3}{32\re{\omega_0}^5\im{\omega_0}}\Biggl(1+\frac{\re{\omega_0}(\re{\omega_0}-\re{\omega_1})}{4\im{\omega_0}^2}
\\\nonumber&&\qquad\qquad +\re{\omega_0}^2\left(\frac{(\re{\omega_0}-\re{\omega_1})^2}{16\im{\omega_0}^4}-\frac{3\im{\omega_0}-\im{\omega_1}}{12\im{\omega_0}}\right)\Biggr)+ \Order{\frac{1}{\kappa^3}},
\end{eqnarray}
where $\omega_1$ denotes the first overtone.

The validity of this correspondence between the quasinormal modes and the S-matrix coefficient $\mathcal{K}$ has been further verified for rotating black holes \cite{Konoplya:2024vuj} and extensively tested in a variety of gravitational backgrounds. Recently, it was shown in \cite{Malik:2025qnr} that Eq.~\eqref{eq:gbsecondorder} can be used to accurately reconstruct the absorption cross section,
\begin{equation}\label{crossection}
\sigma(\omega)=\frac{\pi}{\omega^2}\sum_{\kappa}2\kappa \Gamma(\omega),
\end{equation}
where the grey-body factor $\Gamma(\omega)$ is given by~\eqref{greybodyfactor}.

The correspondence between grey-body factors and quasinormal modes has been extensively examined in numerous recent studies \cite{Antoniou:2025bvg,Pedrotti:2025idg,Malik:2024cgb,Dubinsky:2024vbn,Skvortsova:2024msa,Tang:2025mkk,Lutfuoglu:2025hjy,Lutfuoglu:2025eik,Lutfuoglu:2025kqp,Lutfuoglu:2025blw,Han:2025cal,Malik:2025erb,Dubinsky:2025nxv,Malik:2025qnr,Dubinsky:2025ypj,Dubinsky:2025wns,Lutfuoglu:2025mqa,Yan:2025pvp,Fan:2025ead,Huang:2025rxx,Lutfuoglu:2026fpx,Arbelaez:2026eaz,Han:2026fpn,Malik:2025dxn,Lutfuoglu:2025ldc,Shi:2025gst,Lutfuoglu:2025ohb}. These works generally demonstrate good agreement already for perturbations with multipole number $\ell=s+1$, i.~e., the second lowest allowed multipole number for a field of spin $s$. Even though the expansion is carried out only to second order beyond the eikonal approximation, the relative error typically remains below the percent level and decreases rapidly as $\ell$ increases.

The correspondence is not accurate or even does not work for double-well potentials \cite{Konoplya:2025hgp,Konoplya:2025ixm}, because the WKB formula should be modified for that case. The correspondence was also extended to wormhole spacetimes in \cite{Bolokhov:2024otn}.

\section{Conclusions}\label{sec:conclusions}

In this work we have presented an optimized and extended version of the higher-order WKB Mathematica® package for the computation of quasinormal modes and grey-body factors. While the formalism itself was developed in earlier works, the present implementation introduces two essential improvements: the extension of the WKB expansion up to the sixteenth order developed recently in \cite{Matyjasek:2019eeu} and, more importantly, the replacement of symbolic high-order differentiation by numerical evaluation of derivatives of the effective potential at its maximum.

The latter modification leads to a substantial acceleration of computations, particularly for black-hole spacetimes with complicated analytic structure of the metric function or non-rational effective potentials. As demonstrated in explicit benchmarks, calculations that were previously time-consuming or practically unfeasible at high WKB orders can now be performed within a fraction of a second, while simultaneously reducing precision loss associated with floating-point operations.

In addition to the technical improvements, the package incorporates routines for analytic expansions beyond the leading eikonal approximation, allowing systematic derivation of quasinormal-mode and grey-body-factor expressions in inverse powers of the multipole number. The correspondence between quasinormal modes and grey-body factors, including higher-order corrections, is also implemented, providing a unified framework for analyzing spectral and scattering characteristics within the same computational environment. The limitations of the correspondence is also discussed.

\section*{Author Contributions}
R.A.K.: conceptualization, methodology, software, writing; J.M.: conceptualization, methodology, software, writing; A.Z.: conceptualization, methodology, software, writing. All authors have read and agreed to the published version of the manuscript.

\section*{Funding}
A. Z. was supported by Conselho Nacional de Desenvolvimento Científico e Tecnológico (CNPq).

\section*{Institutional Review Board Statement}
Not applicable.

\section*{Informed Consent Statement}
Not applicable.

\section*{Data Availability Statement}
The Wolfram Mathematica® package and examples of its applications are publicly available online at\\\url{https://arxiv.org/src/2603.12466v1/anc}.

\section*{Conflicts of Interest}
Given their editorial roles, Roman A. Konoplya (Editor-in-Chief) and Alexander Zhidenko (Editorial Board Member) had no involvement in the peer review of this paper and had no access to information regarding its peer-review process. Full responsibility for the editorial process of this paper was delegated to another editor of the journal.

\section*{Use of AI and AI-Assisted Technologies}
During the preparation of this work, the authors used ChatGPT developed by OpenAI for language editing and text refinement. After using this service, the authors reviewed and edited the content as needed and take full responsibility for the content of the published article.

\small
\bibliographystyle{scilight}
\bibliography{bibliography}

@article{Iyer:1986np,
    author = "Iyer, Sai and Will, Clifford M.",
    title = "{Black Hole Normal Modes: A {WKB} Approach. 1. Foundations and Application of a Higher Order {WKB} Analysis of Potential Barrier Scattering}",
    reportNumber = "Print-86-1482 (WASH. U., ST. LOUIS)",
    doi = "10.1103/PhysRevD.35.3621",
    journal = "Phys. Rev. D",
    volume = "35",
    pages = "3621",
    year = "1987"
}

@article{Schutz:1985km,
    author = "Schutz, Bernard F. and Will, Clifford M.",
    title = "{BLACK HOLE NORMAL MODES: A SEMIANALYTIC APPROACH}",
    reportNumber = "PRINT-85-0063 (WASH.U.,ST.LOUIS)",
    doi = "10.1086/184453",
    journal = "Astrophys. J. Lett.",
    volume = "291",
    pages = "L33--L36",
    year = "1985"
}

@article{Konoplya:2003ii,
    author = "Konoplya, R. A.",
    title = "{Quasinormal behavior of the d-dimensional Schwarzschild black hole and higher order WKB approach}",
    eprint = "gr-qc/0303052",
    archivePrefix = "arXiv",
    doi = "10.1103/PhysRevD.68.024018",
    journal = "Phys. Rev. D",
    volume = "68",
    pages = "024018",
    year = "2003"
}

@article{Matyjasek:2017psv,
    author = "Matyjasek, Jerzy and Opala, Michał",
    title = "{Quasinormal modes of black holes. The improved semianalytic approach}",
    eprint = "1704.00361",
    archivePrefix = "arXiv",
    primaryClass = "gr-qc",
    doi = "10.1103/PhysRevD.96.024011",
    journal = "Phys. Rev. D",
    volume = "96",
    number = "2",
    pages = "024011",
    year = "2017"
}

@article{Matyjasek:2019eeu,
    author = "Matyjasek, Jerzy and Telecka, Malgorzata",
    title = "{Quasinormal modes of black holes. II. Padé summation of the higher-order WKB terms}",
    eprint = "1908.09389",
    archivePrefix = "arXiv",
    primaryClass = "gr-qc",
    doi = "10.1103/PhysRevD.100.124006",
    journal = "Phys. Rev. D",
    volume = "100",
    number = "12",
    pages = "124006",
    year = "2019"
}

@article{Konoplya:2019hlu,
    author = "Konoplya, R. A. and Zhidenko, A. and Zinhailo, A. F.",
    title = "{Higher order WKB formula for quasinormal modes and grey-body factors: recipes for quick and accurate calculations}",
    eprint = "1904.10333",
    archivePrefix = "arXiv",
    primaryClass = "gr-qc",
    doi = "10.1088/1361-6382/ab2e25",
    journal = "Class. Quant. Grav.",
    volume = "36",
    pages = "155002",
    year = "2019"
}

@article{Konoplya:2024lir,
    author = "Konoplya, R. A. and Zhidenko, A.",
    title = "{Correspondence between grey-body factors and quasinormal modes}",
    eprint = "2406.11694",
    archivePrefix = "arXiv",
    primaryClass = "gr-qc",
    doi = "10.1088/1475-7516/2024/09/068",
    journal = "JCAP",
    volume = "09",
    pages = "068",
    year = "2024"
}

@article{Konoplya:2024vuj,
    author = "Konoplya, R. A. and Zhidenko, A.",
    title = "{Correspondence between grey-body factors and quasinormal frequencies for rotating black holes}",
    eprint = "2408.11162",
    archivePrefix = "arXiv",
    primaryClass = "gr-qc",
    doi = "10.1016/j.physletb.2025.139288",
    journal = "Phys. Lett. B",
    volume = "861",
    pages = "139288",
    year = "2025"
}

@article{Kokkotas:1999bd,
    author = "Kokkotas, Kostas D. and Schmidt, Bernd G.",
    title = "{Quasinormal modes of stars and black holes}",
    eprint = "gr-qc/9909058",
    archivePrefix = "arXiv",
    doi = "10.12942/lrr-1999-2",
    journal = "Living Rev. Rel.",
    volume = "2",
    pages = "2",
    year = "1999"
}

@article{Konoplya:2011qq,
    author = "Konoplya, R. A. and Zhidenko, A.",
    title = "{Quasinormal modes of black holes: From astrophysics to string theory}",
    eprint = "1102.4014",
    archivePrefix = "arXiv",
    primaryClass = "gr-qc",
    doi = "10.1103/RevModPhys.83.793",
    journal = "Rev. Mod. Phys.",
    volume = "83",
    pages = "793--836",
    year = "2011"
}

@article{Berti:2009kk,
    author = "Berti, Emanuele and Cardoso, Vitor and Starinets, Andrei O.",
    title = "{Quasinormal modes of black holes and black branes}",
    eprint = "0905.2975",
    archivePrefix = "arXiv",
    primaryClass = "gr-qc",
    doi = "10.1088/0264-9381/26/16/163001",
    journal = "Class. Quant. Grav.",
    volume = "26",
    pages = "163001",
    year = "2009"
}

@article{Bolokhov:2025uxz,
    author = "Bolokhov, S. V. and Skvortsova, Milena",
    title = "{Review of analytic results on quasinormal modes of black holes}",
    eprint = "2504.05014",
    doi = "10.1134/S0202289325700306",
    archivePrefix = "arXiv",
    primaryClass = "gr-qc",
    journal = "Grav. Cosmol.",
    volume = "31",
    number = "4",
    pages = "423--446",
    year = "2025"
}

@article{LIGOScientific:2016aoc,
    author = "Abbott, B. P. and others",
    collaboration = "LIGO Scientific, Virgo",
    title = "{Observation of Gravitational Waves from a Binary Black Hole Merger}",
    eprint = "1602.03837",
    archivePrefix = "arXiv",
    primaryClass = "gr-qc",
    reportNumber = "LIGO-P150914",
    doi = "10.1103/PhysRevLett.116.061102",
    journal = "Phys. Rev. Lett.",
    volume = "116",
    number = "6",
    pages = "061102",
    year = "2016"
}

@article{LIGOScientific:2017vwq,
    author = "Abbott, B. P. and others",
    collaboration = "LIGO Scientific, Virgo",
    title = "{GW170817: Observation of Gravitational Waves from a Binary Neutron Star Inspiral}",
    eprint = "1710.05832",
    archivePrefix = "arXiv",
    primaryClass = "gr-qc",
    reportNumber = "LIGO-P170817",
    doi = "10.1103/PhysRevLett.119.161101",
    journal = "Phys. Rev. Lett.",
    volume = "119",
    number = "16",
    pages = "161101",
    year = "2017"
}

@article{LIGOScientific:2020zkf,
    author = "Abbott, R. and others",
    collaboration = "LIGO Scientific, Virgo",
    title = "{GW190814: Gravitational Waves from the Coalescence of a 23 Solar Mass Black Hole with a 2.6 Solar Mass Compact Object}",
    eprint = "2006.12611",
    archivePrefix = "arXiv",
    primaryClass = "astro-ph.HE",
    reportNumber = "LIGO-P190814",
    doi = "10.3847/2041-8213/ab960f",
    journal = "Astrophys. J. Lett.",
    volume = "896",
    number = "2",
    pages = "L44",
    year = "2020"
}

@article{Babak:2017tow,
    author = "Babak, Stanislav and Gair, Jonathan and Sesana, Alberto and Barausse, Enrico and Sopuerta, Carlos F. and Berry, Christopher P. L. and Berti, Emanuele and Amaro-Seoane, Pau and Petiteau, Antoine and Klein, Antoine",
    title = "{Science with the space-based interferometer LISA. V: Extreme mass-ratio inspirals}",
    eprint = "1703.09722",
    archivePrefix = "arXiv",
    primaryClass = "gr-qc",
    doi = "10.1103/PhysRevD.95.103012",
    journal = "Phys. Rev. D",
    volume = "95",
    number = "10",
    pages = "103012",
    year = "2017"
}

@book{Futterman:1988ni,
    author = "Futterman, J. A. H. and Handler, F. A. and Matzner, R. A.",
    title = "{SCATTERING FROM BLACK HOLES}",
    doi = "10.1017/CBO9780511735615",
    isbn = "978-1-139-24539-5, 978-0-521-11210-9",
    publisher = "Cambridge University Press",
    series = "Cambridge Monographs on Mathematical Physics",
    month = "5",
    year = "2012"
}

@article{Page:1976df,
    author = "Page, Don N.",
    title = "{Particle Emission Rates from a Black Hole: Massless Particles from an Uncharged, Nonrotating Hole}",
    doi = "10.1103/PhysRevD.13.198",
    journal = "Phys. Rev. D",
    volume = "13",
    pages = "198--206",
    year = "1976"
}

@article{Konoplya:2023moy,
    author = "Konoplya, R. A. and Zhidenko, A.",
    title = "{Analytic expressions for quasinormal modes and grey-body factors in the eikonal limit and beyond}",
    eprint = "2309.02560",
    archivePrefix = "arXiv",
    primaryClass = "gr-qc",
    doi = "10.1088/1361-6382/ad0a52",
    journal = "Class. Quant. Grav.",
    volume = "40",
    number = "24",
    pages = "245005",
    year = "2023"
}

@article{Bonanno:2025dry,
    author = "Bonanno, Alfio M. and Konoplya, Roman A. and Oglialoro, Giovanni and Spina, Andrea",
    title = "{Regular black holes from proper-time flow in quantum gravity and their quasinormal modes, shadow and Hawking radiation}",
    eprint = "2509.12469",
    archivePrefix = "arXiv",
    primaryClass = "gr-qc",
    doi = "10.1088/1475-7516/2025/12/042",
    journal = "JCAP",
    volume = "12",
    pages = "042",
    year = "2025"
}

@article{Zhidenko:2003wq,
    author = "Zhidenko, A.",
    title = "{Quasinormal modes of Schwarzschild de Sitter black holes}",
    eprint = "gr-qc/0307012",
    archivePrefix = "arXiv",
    doi = "10.1088/0264-9381/21/1/019",
    journal = "Class. Quant. Grav.",
    volume = "21",
    pages = "273--280",
    year = "2004"
}

@article{Arbelaez:2026eaz,
    author = "Arbelaez, Juan Pablo",
    title = "{Grey-body factors of higher dimensional regular black holes in quasi-topological theories}",
    eprint = "2601.22340",
    archivePrefix = "arXiv",
    primaryClass = "gr-qc",
    month = "1",
    year = "2026"
}

@article{Han:2026fpn,
    author = "Han, Hyewon and Gwak, Bogeun",
    title = "{Correspondence between quasinormal modes and grey-body factors of Schwarzschild--Tangherlini black holes}",
    eprint = "2601.18613",
    archivePrefix = "arXiv",
    primaryClass = "gr-qc",
    month = "1",
    year = "2026"
}

@article{Lutfuoglu:2026fpx,
    author = {Lütfüoǧlu, Bekir Can and Rayimbaev, Javlon and Rahmatov, Bekzod and Shayimov, Fayzullo and Davletov, Ikram},
    title = "{Telling tails and quasi-resonances in the vicinity of Dymnikova regular black hole}",
    eprint = "2601.17906",
    archivePrefix = "arXiv",
    primaryClass = "gr-qc",
    month = "1",
    year = "2026"
}

@article{Konoplya:2025hgp,
    author = "Konoplya, R. A. and Stashko, O. S.",
    title = "{Transition from regular black holes to wormholes in covariant effective quantum gravity: Scattering, quasinormal modes, and Hawking radiation}",
    eprint = "2502.05689",
    archivePrefix = "arXiv",
    primaryClass = "gr-qc",
    doi = "10.1103/PhysRevD.111.084031",
    journal = "Phys. Rev. D",
    volume = "111",
    number = "8",
    pages = "084031",
    year = "2025"
}

@article{Lutfuoglu:2025hjy,
    author = {Lütfüoǧlu, B. C.},
    title = "{Long-lived quasinormal modes and gray-body factors of black holes and wormholes in dark matter inspired Weyl gravity}",
    eprint = "2503.16087",
    archivePrefix = "arXiv",
    primaryClass = "gr-qc",
    doi = "10.1140/epjc/s10052-025-14210-0",
    journal = "Eur. Phys. J. C",
    volume = "85",
    number = "5",
    pages = "486",
    year = "2025"
}

@article{Tang:2025mkk,
    author = "Tang, Chen and Ling, Yi and Jiang, Qing-Quan",
    title = "{Correspondence between grey-body factors and quasinormal modes for regular black holes with sub-Planckian curvature*}",
    eprint = "2503.21597",
    archivePrefix = "arXiv",
    primaryClass = "gr-qc",
    doi = "10.1088/1674-1137/adfa74",
    journal = "Chin. Phys.",
    volume = "49",
    number = "12",
    pages = "125110",
    year = "2025"
}

@article{Huang:2025rxx,
    author = "Huang, Zun-Xian and Li, Peng-Cheng",
    title = "{Quasinormal mode/grey-body factor correspondence for Kerr black holes}",
    eprint = "2512.23510",
    archivePrefix = "arXiv",
    primaryClass = "gr-qc",
    month = "12",
    year = "2025"
}

@article{Pedrotti:2025idg,
    author = "Pedrotti, Davide and Calzà, Marco",
    title = "{Trinity of black hole correspondences: Shadows, quasinormal modes, graybody factors, and cautionary remarks}",
    eprint = "2504.01909",
    archivePrefix = "arXiv",
    primaryClass = "gr-qc",
    doi = "10.1103/1q35-mjjz",
    journal = "Phys. Rev. D",
    volume = "111",
    number = "12",
    pages = "124056",
    year = "2025"
}

@article{Lutfuoglu:2025mqa,
    author = {Lütfüoǧlu, Bekir Can and Shermatov, Abubakir and Rayimbaev, Javlon and Matyoqubov, Muhammad and Sirajiddin, Otaboyev},
    title = "{Gravitational spectra and wave propagation in regular black holes supported by a Dehnen Halo}",
    eprint = "2511.22366",
    archivePrefix = "arXiv",
    primaryClass = "gr-qc",
    doi = "10.1140/epjc/s10052-025-15234-2",
    journal = "Eur. Phys. J. C",
    volume = "85",
    number = "12",
    pages = "1484",
    year = "2025"
}

@article{Lutfuoglu:2025ohb,
    author = {Lütfüoǧlu, B. C.},
    title = "{Quasinormal modes and gray-body factors for gravitational perturbations in asymptotically safe gravity}",
    eprint = "2505.06966",
    archivePrefix = "arXiv",
    primaryClass = "gr-qc",
    doi = "10.1140/epjc/s10052-026-15290-2",
    journal = "Eur. Phys. J. C",
    volume = "86",
    number = "1",
    pages = "39",
    year = "2026"
}

@article{Shi:2025gst,
    author = "Shi, Qi-Long and Wang, Rui and Xiong, Wei and Li, Peng-Cheng",
    title = "{Quasinormal modes and grey-body factors of axial gravitational perturbations of regular black holes in asymptotically safe gravity}",
    eprint = "2506.16217",
    archivePrefix = "arXiv",
    primaryClass = "gr-qc",
    month = "6",
    year = "2025"
}

@article{Yan:2025pvp,
    author = "Yan, Hao-Peng and Zhang, Zeng-Yi and Yue, Xiao-Jun and Li, Xiang-Qian",
    title = "{Black holes immersed in modified Chaplygin-like dark fluid and cloud of strings: shadows, quasinormal modes, and greybody factors*}",
    eprint = "2511.21205",
    archivePrefix = "arXiv",
    primaryClass = "gr-qc",
    doi = "10.1088/1674-1137/ae30ea",
    journal = "Chin. Phys.",
    volume = "50",
    number = "4",
    pages = "045102",
    year = "2026"
}

@article{Konoplya:2025ixm,
    author = "Konoplya, Roman A. and Pappas, Thomas D.",
    title = "{Dirty black holes, clean signals: near-horizon vs.~environmental effects on grey-body factors and Hawking radiation}",
    eprint = "2507.01954",
    archivePrefix = "arXiv",
    primaryClass = "gr-qc",
    doi = "10.1088/1475-7516/2026/02/038",
    journal = "JCAP",
    volume = "02",
    pages = "038",
    year = "2026"
}

@article{Skvortsova:2024msa,
    author = "Skvortsova, Milena",
    title = "{Quantum corrected black holes: testing the correspondence between grey-body factors and quasinormal modes}",
    eprint = "2411.06007",
    archivePrefix = "arXiv",
    primaryClass = "gr-qc",
    doi = "10.1140/epjc/s10052-025-14589-w",
    journal = "Eur. Phys. J. C",
    volume = "85",
    number = "8",
    pages = "854",
    year = "2025"
}

@article{Dubinsky:2024vbn,
    author = "Dubinsky, Alexey",
    title = "{Gray-body factors for gravitational and electromagnetic perturbations around Gibbons-Maeda-Garfinkle-Horowitz-Strominger black holes}",
    eprint = "2412.00625",
    archivePrefix = "arXiv",
    primaryClass = "gr-qc",
    doi = "10.1142/S0217732325501111",
    journal = "Mod. Phys. Lett. A",
    volume = "40",
    number = "28",
    pages = "2550111",
    year = "2025"
}

@article{Bolokhov:2024otn,
    author = "Bolokhov, S. V. and Skvortsova, Milena",
    title = "{Correspondence between quasinormal modes and grey-body factors of spherically symmetric traversable wormholes}",
    eprint = "2412.11166",
    archivePrefix = "arXiv",
    primaryClass = "gr-qc",
    doi = "10.1088/1475-7516/2025/04/025",
    journal = "JCAP",
    volume = "04",
    pages = "025",
    year = "2025"
}

@article{Malik:2024cgb,
    author = "Malik, Zainab",
    title = "{Correspondence between quasinormal modes and grey-body factors for massive fields in Schwarzschild-de~Sitter spacetime}",
    eprint = "2412.19443",
    archivePrefix = "arXiv",
    primaryClass = "gr-qc",
    doi = "10.1088/1475-7516/2025/04/042",
    journal = "JCAP",
    volume = "04",
    pages = "042",
    year = "2025"
}

@article{Lutfuoglu:2025ldc,
    author = {Lütfüoǧlu, Bekir Can},
    title = "{Black Holes in Proca-Gauss-Bonnet Gravity with Primary Hair: Particle Motion, Shadows, and Grey-Body Factors}",
    eprint = "2507.09246",
    archivePrefix = "arXiv",
    primaryClass = "gr-qc",
    doi = "10.53941/ijgtp.2025.100004",
    journal = "Int. J. Grav. Theor. Phys.",
    volume = "1",
    number = "1",
    pages = "4",
    year = "2025"
}

@article{Antoniou:2025bvg,
    author = "Antoniou, Georgios and Pappas, Thomas D. and Kanti, Panagiota",
    title = "{Greybody factors in scalar-tensor gravity and beyond}",
    eprint = "2507.17329",
    archivePrefix = "arXiv",
    primaryClass = "gr-qc",
    doi = "10.1103/zwhl-sqqs",
    journal = "Phys. Rev. D",
    volume = "112",
    number = "8",
    pages = "084013",
    year = "2025"
}

@article{Fan:2025ead,
    author = "Fan, SiHao and Wu, Chen and Guo, WenJun",
    title = "{Grey-body Factors and Absorption Cross Sections of Non-Commutative Black Holes under Einstein-Coupled Scalar Fields}",
    eprint = "2511.16012",
    archivePrefix = "arXiv",
    primaryClass = "gr-qc",
    month = "11",
    year = "2025"
}

@article{Han:2025cal,
    author = "Han, Hyewon and Gwak, Bogeun",
    title = "{Correspondence between quasinormal modes and grey-body factors in five-dimensional black holes}",
    eprint = "2508.12989",
    archivePrefix = "arXiv",
    primaryClass = "gr-qc",
    month = "8",
    year = "2025"
}

@article{Malik:2025dxn,
    author = "Malik, Zainab",
    title = "{Gravitational Perturbations of the Hayward Spacetime and Testing the Correspondence between Quasinormal Modes and Grey-body Factors}",
    eprint = "2508.19178",
    archivePrefix = "arXiv",
    primaryClass = "gr-qc",
    doi = "10.1007/s10773-025-06198-w",
    journal = "Int. J. Theor. Phys.",
    volume = "64",
    number = "11",
    pages = "314",
    year = "2025"
}

@article{Dubinsky:2025wns,
    author = "Dubinsky, Alexey",
    title = "{Long-lived Modes and Grey-body Factors of Massive Fields in Quantum-corrected (Hayward) Black Holes}",
    eprint = "2511.00778",
    archivePrefix = "arXiv",
    primaryClass = "gr-qc",
    doi = "10.1007/s10773-026-06274-9",
    journal = "Int. J. Theor. Phys.",
    volume = "65",
    number = "2",
    pages = "45",
    year = "2026"
}

@article{Dubinsky:2025nxv,
    author = "Dubinsky, Alexey",
    title = "{Gravitational perturbations of Dymnikova black holes: Grey-body factors and absorption cross-sections}",
    eprint = "2509.11017",
    archivePrefix = "arXiv",
    primaryClass = "gr-qc",
    doi = "10.1016/j.aop.2025.170299",
    journal = "Annals Phys.",
    volume = "485",
    pages = "170299",
    year = "2026"
}

@article{Malik:2025erb,
    author = "Malik, Zainab",
    title = "{Grey-Body Factors for Scalar and Dirac Fields in the Euler-Heisenberg Electrodynamics}",
    eprint = "2509.15995",
    archivePrefix = "arXiv",
    primaryClass = "gr-qc",
    doi = "10.53941/ijgtp.2025.100006",
    journal = "Int. J. Grav. Theor. Phys.",
    volume = "1",
    number = "1",
    pages = "6",
    year = "2025"
}

@article{Lutfuoglu:2025blw,
    author = {Lütfüoǧlu, Bekir Can and Saka, Erdinç Ulaş and Shermatov, Abubakir and Rayimbaev, Javlon and Ibragimov, Inomjon and Muminov, Sokhibjan},
    title = "{Proper-time approach in asymptotic safety via black hole quasinormal modes and grey-body factors}",
    eprint = "2509.15923",
    archivePrefix = "arXiv",
    primaryClass = "gr-qc",
    doi = "10.1140/epjc/s10052-025-14950-z",
    journal = "Eur. Phys. J. C",
    volume = "85",
    number = "10",
    pages = "1190",
    year = "2025"
}

@article{Malik:2025qnr,
    author = "Malik, Zainab",
    title = "{Bonanno-Reuter regular black hole: quasi-resonances, grey-body factors and absorption cross-sections of a massive scalar field}",
    eprint = "2510.06689",
    archivePrefix = "arXiv",
    primaryClass = "gr-qc",
    month = "10",
    year = "2025"
}

@article{Lutfuoglu:2025eik,
    author = {Lütfüoǧlu, Bekir Can},
    title = "{Grey-Body Factors and Absorption Cross-Sections~of Scalar and Dirac Fields in the Vicinity of Dilaton-De Sitter Black Hole}",
    eprint = "2510.10579",
    archivePrefix = "arXiv",
    primaryClass = "gr-qc",
    doi = "10.1002/prop.70074",
    journal = "Fortsch. Phys.",
    volume = "74",
    number = "1",
    pages = "e70074",
    year = "2026"
}

@article{Dubinsky:2025ypj,
    author = "Dubinsky, Alexey",
    title = "{Scattering and Absorption of Standard Model Fields by Brane-Localized Schwarzschild--de Sitter Black Holes}",
    eprint = "2510.11643",
    archivePrefix = "arXiv",
    primaryClass = "gr-qc",
    month = "10",
    year = "2025"
}

@article{Lutfuoglu:2025kqp,
    author = {Lütfüoǧlu, B. C.},
    title = "{Long-lived quasinormal modes, grey-body factors and absorption cross-section of the black hole immersed in the Hernquist galactic halo}",
    eprint = "2510.25969",
    archivePrefix = "arXiv",
    primaryClass = "gr-qc",
    doi = "10.1016/j.physletb.2025.140082",
    journal = "Phys. Lett. B",
    volume = "872",
    pages = "140082",
    year = "2026"
}

@article{Malik:2024qsz,
    author = "Malik, Zainab",
    title = "{Analytic quasinormal frequencies of the regular Simpson-Visser black hole}",
    doi = "10.1142/S0217751X2450132X",
    journal = "Int. J. Mod. Phys. A",
    volume = "40",
    number = "02",
    pages = "2450132",
    year = "2025"
}

@article{Malik:2024bmp,
    author = "Malik, Zainab",
    title = "{Quasinormal modes of the Mannheim-Kazanas black holes}",
    doi = "10.1515/zna-2024-0153",
    journal = "Z. Naturforsch. A",
    volume = "79",
    number = "11",
    pages = "1063--1073",
    year = "2024"
}

@article{Malik:2024itg,
    author = "Malik, Zainab",
    title = "{Analytic quasinormal frequencies of the Casadio-Fabbri-Mazzacurati black hole}",
    doi = "10.1139/cjp-2024-0247",
    journal = "Can. J. Phys.",
    volume = "103",
    number = "9",
    pages = "800--811",
    year = "2025"
}

@article{Malik:2024tuf,
    author = "Malik, Zainab",
    title = "{Analytical QNMs of fields of various spin in the Hayward spacetime}",
    eprint = "2410.04306",
    archivePrefix = "arXiv",
    primaryClass = "gr-qc",
    reportNumber = "Research Gate Preprint: DOI:10.13140/RG.2.2.32496.06402",
    doi = "10.1209/0295-5075/ad7885",
    journal = "EPL",
    volume = "147",
    number = "6",
    pages = "69001",
    year = "2024"
}

@article{Bolokhov:2023bwm,
    author = "Bolokhov, S. V.",
    title = "{Long-lived quasinormal modes and overtones’ behavior of holonomy-corrected black holes}",
    eprint = "2311.05503",
    archivePrefix = "arXiv",
    primaryClass = "gr-qc",
    doi = "10.1103/PhysRevD.110.024010",
    journal = "Phys. Rev. D",
    volume = "110",
    number = "2",
    pages = "024010",
    year = "2024"
}

@article{Malik:2024wvs,
    author = "Malik, Zainab",
    title = "{Quasinormal modes and grey-body factors of Morris-Thorne wormholes}",
    eprint = "2412.13385",
    archivePrefix = "arXiv",
    primaryClass = "gr-qc",
    month = "12",
    year = "2024"
}

@article{Konoplya:2025mvj,
    author = "Konoplya, Roman A. and Khrabustovskyi, Andrii and Kříž, Jan and Zhidenko, Alexander",
    title = "{Quasinormal ringing and shadows of black holes and wormholes in dark matter-inspired Weyl gravity}",
    eprint = "2501.16134",
    archivePrefix = "arXiv",
    primaryClass = "gr-qc",
    doi = "10.1088/1475-7516/2025/04/062",
    journal = "JCAP",
    volume = "04",
    pages = "062",
    year = "2025"
}

@article{Sekhmani:2025jbl,
    author = "Sekhmani, Y. and Baruah, A. and Maurya, S. K. and Rayimbaev, J. and Altanji, M. and Ibragimov, I. and Muminov, S.",
    title = "{Kalb-Ramond black holes sourced by ModMax electrodynamics: Some perturbative properties in the phantom sector}",
    eprint = "2507.19088",
    archivePrefix = "arXiv",
    primaryClass = "gr-qc",
    doi = "10.1016/j.dark.2025.102157",
    journal = "Phys. Dark Univ.",
    volume = "50",
    pages = "102157",
    year = "2025"
}

@article{Ou:2025bbv,
    author = "Ou, Minyan and Zhang, Xiangdong",
    title = "{Quantum Oppenheimer-Snyder models in loop quantum cosmology with Lorentz term}",
    eprint = "2508.01183",
    archivePrefix = "arXiv",
    primaryClass = "gr-qc",
    doi = "10.1103/d4fp-nwbf",
    journal = "Phys. Rev. D",
    volume = "112",
    number = "12",
    pages = "126016",
    year = "2025"
}

@article{Lutfuoglu:2025qkt,
    author = {Lütfüoǧlu, B. C.},
    title = "{Long-lived quasinormal modes and echoes in the Einstein-Gauss-Bonnet-Proca theory}",
    eprint = "2508.19194",
    archivePrefix = "arXiv",
    primaryClass = "gr-qc",
    doi = "10.1140/epjc/s10052-025-14839-x",
    journal = "Eur. Phys. J. C",
    volume = "85",
    number = "9",
    pages = "1076",
    year = "2025"
}

@article{Bolokhov:2025egl,
    author = "Bolokhov, S. V. and Skvortsova, Milena",
    title = "{Gravitational quasinormal modes of the Hayward spacetime}",
    eprint = "2508.19989",
    archivePrefix = "arXiv",
    primaryClass = "gr-qc",
    month = "8",
    year = "2025"
}

@article{Deng:2025hfn,
    author = "Deng, Weike and Long, Sheng and Tan, Qin and Chen, Zu-Cheng and Jing, Jiliang",
    title = "{Scalar-gravitational quasinormal modes and echoes in a five dimensional thick brane}",
    eprint = "2508.20937",
    archivePrefix = "arXiv",
    primaryClass = "gr-qc",
    doi = "10.1007/JHEP01(2026)066",
    journal = "JHEP",
    volume = "01",
    pages = "066",
    year = "2026"
}

@article{Bambagiotti:2025qxj,
    author = "Bambagiotti, Tommaso and Gallerani, Luca and Mentrelli, Andrea and Giusti, Andrea and Casadio, Roberto",
    title = "{Quantum dust cores of black holes and their quasi-normal modes}",
    eprint = "2509.01570",
    archivePrefix = "arXiv",
    primaryClass = "gr-qc",
    month = "9",
    year = "2025"
}

@article{Feng:2025iao,
    author = "Feng, Xing-Hui and Zhang, Guang-Yu",
    title = "{Shadow and quasi-normal modes of Schwarzschild-Hernquist black hole}",
    eprint = "2509.04001",
    archivePrefix = "arXiv",
    primaryClass = "gr-qc",
    doi = "10.1140/epjc/s10052-026-15293-z",
    journal = "Eur. Phys. J. C",
    volume = "86",
    number = "1",
    pages = "36",
    year = "2026"
}

@article{AraujoFilho:2025vgb,
    author = "Araújo Filho, A. A. and Heidari, N. and Lobo, Iarley P.",
    title = "{Black Hole Gravitational Phenomena in Higher-Order Curvature-Scalar Gravity}",
    eprint = "2509.11985",
    archivePrefix = "arXiv",
    primaryClass = "gr-qc",
    month = "9",
    year = "2025"
}

@article{Liu:2025iby,
    author = "Liu, Yunlong and Zhang, Xiangdong",
    title = "{Quasinormal modes and tidal Love numbers of covariant effective quantum black holes with cosmological constant}",
    eprint = "2509.12013",
    archivePrefix = "arXiv",
    primaryClass = "gr-qc",
    month = "9",
    year = "2025"
}

@article{Aoki:2025ihc,
    author = "Aoki, Katsuki and Cristofoli, Andrea and Jeong, Hyun and Sergola, Matteo and Yoshimura, Kaho",
    title = "{Quantum effects for black holes with on-shell amplitudes}",
    eprint = "2509.12111",
    archivePrefix = "arXiv",
    primaryClass = "hep-th",
    reportNumber = "YITP-24-143, UT-Komaba/25-9, IPMU25-0045, RESCEU-19/25",
    doi = "10.1007/JHEP12(2025)163",
    journal = "JHEP",
    volume = "12",
    pages = "163",
    year = "2025"
}

@article{Bohra:2025oro,
    author = "Bohra, Sunil Singh",
    title = "{Blackhole perturbations in the Modified Generalized Chaplygin Gas model}",
    eprint = "2509.14064",
    archivePrefix = "arXiv",
    primaryClass = "gr-qc",
    month = "9",
    year = "2025"
}

@article{Arbelaez:2025gwj,
    author = "Arbelaez, Juan Pablo",
    title = "{Quasinormal spectra of higher dimensional regular black holes in theories with infinite curvature corrections}",
    eprint = "2509.25141",
    archivePrefix = "arXiv",
    primaryClass = "gr-qc",
    month = "9",
    year = "2025"
}

@article{Lutfuoglu:2025pzi,
    author = {Lütfüoǧlu, Bekir Can and Saka, Erdinç Ulaş and Shermatov, Abubakir and Rayimbaev, Javlon and Ibragimov, Inomjon and Muminov, Sokhibjan},
    title = "{Gravitational quasinormal modes of Dymnikova black holes}",
    eprint = "2509.24633",
    archivePrefix = "arXiv",
    primaryClass = "gr-qc",
    doi = "10.1016/j.aop.2026.170360",
    journal = "Annals Phys.",
    volume = "487",
    pages = "170360",
    year = "2026"
}

@article{Hamil:2025fbn,
    author = "Hamil, B. and Birkandan, T.",
    title = "{4D Einstein-Gauss-Bonnet Black Holes Surrounded by Quintessence in Noncommutative Spacetime}",
    doi = "10.1016/j.nuclphysb.2025.117145",
    journal = "Nucl. Phys. B",
    volume = "1020",
    pages = "117145",
    year = "2025"
}

@article{Pathrikar:2025gzu,
    author = "Pathrikar, Akshat",
    title = "{Quasinormal Ringing and Unruh-Verlinde Temperature of the Frolov Black Hole}",
    eprint = "2510.01376",
    archivePrefix = "arXiv",
    primaryClass = "gr-qc",
    doi = "10.53941/ijgtp.2026.100001",
    journal = "Int. J. Grav. Theor. Phys.",
    volume = "1",
    pages = "1",
    year = "2026"
}

@article{Sajjad:2025xdo,
    author = "Sajjad, W. and Azam, M.",
    title = "{Study of Kiselev black hole in quantum fluctuation modified gravity via quasinormal modes and greybody factors}",
    eprint = "2510.02409",
    archivePrefix = "arXiv",
    primaryClass = "gr-qc",
    doi = "10.1088/1402-4896/ae43d6",
    journal = "Phys. Scripta",
    volume = "101",
    number = "9",
    pages = "095301",
    year = "2026"
}

@article{Momennia:2025vkv,
    author = "Momennia, Mehrab",
    title = {{Ringing of Reissner-Nordström Black Holes with a Non-Abelian Hair in Gravity’s Rainbow}},
    doi = "10.3390/universe11100341",
    journal = "Universe",
    volume = "11",
    number = "10",
    pages = "341",
    year = "2025"
}

@article{Spina:2025wxb,
    author = "Spina, Andrea",
    title = "{Black Holes in Asymptotic Safety: A Review of Solutions and Phenomenology}",
    eprint = "2510.14552",
    archivePrefix = "arXiv",
    primaryClass = "gr-qc",
    doi = "10.53941/ijgtp.2025.100008",
    journal = "Int. J. Grav. Theor. Phys.",
    volume = "1",
    number = "1",
    pages = "8",
    year = "2025"
}

@article{Liang:2025rbe,
    author = "Liang, Qi-Qi and Cai, Ziqiang and Liu, Dong and Long, Zheng-Wen",
    title = "{Observational properties and quasinormal Modes of the Hayward black Hole surrounded by a cloud of strings}",
    eprint = "2511.02396",
    archivePrefix = "arXiv",
    primaryClass = "gr-qc",
    month = "11",
    year = "2025"
}

@article{Pathrikar:2025sin,
    author = "Pathrikar, Akshat",
    title = "{Signatures of a Schwarzschild-like Black Hole Immersed in Dark Matter Halo}",
    eprint = "2511.02355",
    archivePrefix = "arXiv",
    primaryClass = "gr-qc",
    month = "11",
    year = "2025"
}

@article{Malik:2025czt,
    author = "Malik, Zainab",
    title = "{Long-lived quasinormal modes and grey-body factors of supermassive black holes with a dark matter halo}",
    eprint = "2511.12335",
    archivePrefix = "arXiv",
    primaryClass = "gr-qc",
    month = "11",
    year = "2025"
}

@article{AraujoFilho:2025zaj,
    author = "Araújo Filho, A. A. and Heidari, N. and Lobo, Iarley P. and Bezerra, V. B.",
    title = "{Gravitational aspects of a new bumblebee black hole}",
    eprint = "2511.12839",
    archivePrefix = "arXiv",
    primaryClass = "gr-qc",
    month = "11",
    year = "2025"
}

@article{Bolokhov:2025fto,
    author = "Bolokhov, S. V.",
    title = "{Quasinormal ringing of a regular black hole sourced by the Dehnen-type distribution of matter}",
    eprint = "2511.12859",
    archivePrefix = "arXiv",
    primaryClass = "gr-qc",
    doi = "10.1016/j.aop.2026.170416",
    journal = "Annals Phys.",
    volume = "488",
    pages = "170416",
    year = "2026"
}

@article{Sekhmani:2025yjf,
    author = "Sekhmani, Y. and Maurya, S. K. and Rayimbaev, J. and Ali, Akram and Jasim, M. K. and Ibragimov, I. and Muminov, S.",
    title = "{Black holes surrounded by a Murnaghan-fluid scalar gas in a global-monopole background}",
    doi = "10.1016/j.dark.2025.102176",
    journal = "Phys. Dark Univ.",
    volume = "50",
    pages = "102176",
    year = "2025"
}

@article{Bolokhov:2025zva,
    author = "Bolokhov, S. V.",
    title = "{Revisiting black holes in dark-matter halos: on consistent solutions to the Einstein equations}",
    eprint = "2512.06930",
    archivePrefix = "arXiv",
    primaryClass = "gr-qc",
    month = "12",
    year = "2025"
}

@article{dePaiva:2025eux,
    author = "de Paiva, R. C. and Alves, K. S. and Cavalcanti, R. T. and da Rocha, R.",
    title = "{Gravitational decoupling and regular hairy black holes: Geodesic stability, quasinormal modes, and thermodynamic properties}",
    eprint = "2512.14920",
    archivePrefix = "arXiv",
    primaryClass = "gr-qc",
    month = "12",
    year = "2025"
}

@article{Tang:2025qaq,
    author = {Tang, Ruijing and Franchini, Nicola and Völkel, Sebastian H. and Berti, Emanuele},
    title = "{Quasinormal modes of rotating black holes beyond general relativity in the WKB approximation}",
    eprint = "2512.17786",
    archivePrefix = "arXiv",
    primaryClass = "gr-qc",
    month = "12",
    year = "2025"
}

@article{Liang:2026eic,
    author = "Liang, Jie and Liu, Dong and Long, Zheng-Wen",
    title = "{Quasinormal modes and greybody factors of black holes corrected by nonlinear electrodynamics}",
    doi = "10.1140/epjc/s10052-025-15245-z",
    journal = "Eur. Phys. J. C",
    volume = "86",
    number = "1",
    pages = "17",
    year = "2026"
}

@article{Pereira:2026ffn,
    author = "Pereira, C. F. S. and Belich, H. and Soares, A. R. and Silva, Marcos V. de S. and Vitória, R. L. L. and Araújo Filho, A. A.",
    title = "{Light propagation and quasinormal modes of a topologically charged Schwarzschild-Klinkhamer wormhole}",
    eprint = "2601.16305",
    archivePrefix = "arXiv",
    primaryClass = "gr-qc",
    month = "1",
    year = "2026"
}

@article{Qi:2026zrr,
    author = "Qi, Shitao and Cai, Ziqiang",
    title = "{Shadows and quasinormal modes of a Schwarzschild black hole immersed in Hernquist dark matter halo}",
    doi = "10.1140/epjc/s10052-026-15331-w",
    journal = "Eur. Phys. J. C",
    volume = "86",
    number = "1",
    pages = "94",
    year = "2026"
}

@article{Lutfuoglu:2026xlo,
    author = {Lütfüoǧlu, Bekir Can and Murodov, Sardor and Abdullaev, Mardon and Rayimbaev, Javlon and Akhmedov, Munisbek and Matyoqubov, Muhammad},
    title = "{Two types of quasinormal modes of Casadio-Fabbri-Mazzacurati brane-world black holes}",
    eprint = "2602.11001",
    archivePrefix = "arXiv",
    primaryClass = "gr-qc",
    month = "2",
    year = "2026"
}

@article{Hatsuda:2021gtn,
    author = "Hatsuda, Yasuyuki and Kimura, Masashi",
    title = "{Spectral Problems for Quasinormal Modes of Black Holes}",
    eprint = "2111.15197",
    archivePrefix = "arXiv",
    primaryClass = "gr-qc",
    reportNumber = "RUP-21-22",
    doi = "10.3390/universe7120476",
    journal = "Universe",
    volume = "7",
    number = "12",
    pages = "476",
    year = "2021"
}
	
\end{document}